\documentclass{article}
\usepackage{lipsum}
\usepackage{authblk}
\usepackage[english]{babel}
\usepackage{amsfonts}
\usepackage{amsmath}
\usepackage[top=2cm, bottom=2cm, left=2cm, right=2cm]{geometry}
\usepackage{fancyhdr}
\usepackage{multicol}
\usepackage{epsfig}
\usepackage{graphics}
\usepackage{epsfig}
\usepackage{lipsum}
\usepackage{fancyhdr}
\usepackage{cite}
\usepackage[table]{xcolor}
\usepackage{multirow}
\renewenvironment{abstract}{
	
	\hfill\begin{minipage}{0.95\textwidth}
		\rule{\textwidth}{1pt}}
	{\par\noindent\rule{\textwidth}{1pt}\end{minipage}
}


\begin{document}
	%
	\title{\textbf{Dual Non-local Controlled-not Gate}}
	\author[1,2]{ \textbf{C. Seida}}
	\author[1]{ \textbf{A. El Allati}}
	\author[1]{ \textbf{K. El Anouz}}
	\affil[1]{\small Laboratory of R\&D in Engineering Sciences, Faculty of Sciences and Techniques Al-Hoceima, Abdelmalek Essaadi University, Tetouan,
		Morocco}
	\affil[2]{\small ESMaR, Mohammed V University, Faculty of Sciences in Rabat, Morocco}
	
	\maketitle
	\begin{abstract}
		
Distant quantum control via quantum gates represents an essential step toward the realization of distributed quantum networks. In this regard, an efficient theoretical protocol for the dual non-local implementation of controlled-not (CNOT) gates between two separated partners is presented. The suggested protocol requires 1~ebit with local operations and classical communication channels. The efficiency of the teleportation scheme is quantified through an infidelity measure. The numerical results show that the infidelity of performing the CNOT gate  between legitimate partners depends on the initial qubit settings. It is also shown that the protocol is performed with high efficiency if the CNOT control qubit and the auxiliary qubit are prepared in the same direction. Furthermore, we provide a noise analysis for the suggested scheme. We find that by maintaining the noise strengths under the threshold $\frac{1}{4}$, one can achieve the dual non-local CNOT gate optimally.
	\end{abstract}
\\


\section{Introduction}\label{section1}
A quantum computer has the ability to efficiently solve the most sophisticated tasks that are intractable on a conventional computer \cite{0}. The most obvious examples are speeding up an unstructured search \cite{1} and factorizing huge numbers \cite{2,2a}. However, these kinds of complicated problems necessitate the construction of a quantum computer with a large seize that performs locally more than $50$ qubits \cite{3}. Moreover, processing procedures for a large number of qubits  with robust efficiency need to prevent their interaction with each other, besides preserving them from their surrounding environments \cite{4}. However, an alternative way to do so is by considering a network of quantum computers where each quantum computer contains only a small number of qubits. Importantly, it has been shown that in quantum networks, many quantum computers can be connected to each other through quantum teleportation (QT) \cite{5}. The latest protocol is a process that allows the transmission of a quantum state from one location to another without physically moving the quantum system itself \cite{5a}.  The QT process was suggested in $1993$ by $Bennet$ $et$ $al$. \cite {6}  as an alternative way to directly transmit information between two legitimate partners due to the high sensitivity of qubits \cite{7}. In light of its importance, it has become a research hot-spot in the last two decades. Nevertheless, there are plenty of modified teleportation protocols, such as controlled teleportation \cite{8,9}, bidirectional quantum teleportation (BQT) \cite{10,11,12}, cyclic teleportation \cite{13} and multi-party quantum teleportation \cite{14}. As a matter of fact, QT is achieved experimentally by means of different substrates, such as photonic qubits \cite{23,24}, solid states \cite{25}, high-dimensional photonic quantum states \cite{25a}, infinite dimensional quantum states \cite{25b} trapped atoms \cite{26}, and atomic states \cite{27}.\\

Most importantly, QT was not only limited to quantum states, but it was extended to quantum gates \cite{3,15,16,17}, which are basic operations that manipulate the state of a quantum system \cite{18}. Quantum gates are the fundamental building blocks of quantum algorithms \cite{19}, and they are used to perform a wide range of tasks, including quantum state preparation \cite{20}, quantum state measurement \cite{21}, and quantum state manipulation \cite{22}. Actually, quantum gate teleportation (QGT) is a technique to perform local gate operations between spatially separated quantum computers, so that it can be used to establish links among distributed quantum computing networks \cite{28,29}. The main issue of quantum gate teleportation is discussed by $Eisert$ $et$ $al.$ \cite{3} where they have presented a protocol that implements a Controlled-Not~(CNOT) gate \cite{18} using only one $ebit$ and one classical bit. In addition, $Chou$ $et$ $al.$ \cite{y} have experimentally realized  a deterministic teleportation of $CNOT$ gate  by utilizing real-time adaptive control. Moreover,  $Wan$ $et$ $al.$  \cite{x} have demonstrated a deterministic QGT  between two qubits in spatially separated locations in an ion trap. $Huang$ $et$ $al.$ \cite{30} have reported an experimental demonstration of teleportation of  controlled-NOT gates assisted with linear optical manipulations. However, $Daiss$ $et$ $al.$ have suggested  a non-local quantum gate implementation between two distant quantum modules that are connected by a $60$ meter fiber link \cite{31}. Indeed, all these teleportation protocols are unidirectional, meaning that the gate implementation can be assured only in one direction from a controller to a target, namely from Alice to Bob. However, a significant step in the path towards a scalable quantum network is making the quantum teleportation of quantum operations possible in two opposite directions, i.e., from Alice to Bob and inversely from Bob to Alice. \\

In this paper, we present a controlled scheme for two-way non-local  implementation of the CNOT gate between separated qubits that are in the possession of Alice and Bob (Fig.1). In the suggested scheme, one needs only the minimal quantum cost, i.e., $1$ ebit, namely a Bell state $|\psi^{+}_{AB}\rangle = \frac{1}{\sqrt{2}}(|00\rangle + |11\rangle)$,  along with local operations and classical communications (LOCC) \cite{32}. We specifically investigated the efficiency of the suggested scheme using the infidelity measure, which is considered as an error quantifier of the teleportation scheme in our context. Moreover, we discuss the effect of the initial state settings and the auxiliary qubit on the dual non-local CNOT gate. Our results show that the optimal implementation of the dual non-local CNOT gate can be achieved if the initial qubit and the auxiliary qubit are prepared in the same direction. Finally, we provide a noise analysis for the suggested protocol, by considering a realistic situation in which the teleportation quantum channel is affected by a generalized amplitude damping channel.
\\

The rest of the present paper is organized as follows: In Sec.2, we introduce the suggested protocol. Also, the implementation of the quantum teleportation circuit is given. Moreover,  we evaluate the infidelity as a  figure of merit to assess our scheme. In Sec.3, we present a noise analysis with a detailed comparison between the present and the previous schemes. Finally,  we summarize our results in Sec.4.

\begin{figure}[!htb]
	\begin{center}
		\includegraphics[scale=1]{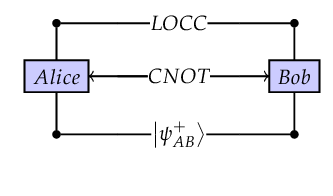}
		\caption{A sketch to show the idea of dual CNOT gate. Two partners Alice and Bob share a Bell state $|\psi^{+}_{AB}\rangle$ and classical channels with local operations (LOCC).}
		\label{FIGURE0}
	\end{center}
\end{figure}

\section{The suggested protocol}
\subsection{Dual teleportation of CNOT gate}
In this section, we describe a scheme to perform a dual non-local CNOT gate between two qubits, one of them at Alice's hand and the other at Bob's hand. The initial states at the partners' sides  are necessarily arbitrary, namely, Alice has the state:
\begin{equation}
	|\Psi_A\rangle=\Big(\cos(\frac{\Theta_A}{2})|0\rangle+e^{i\phi_A}\sin(\frac{\Theta_A}{2})|1\rangle\Big)_{a},
\end{equation}
Similarly, Bob has the following state:
\begin{equation}
	|\Psi_B\rangle=\Big(\cos(\frac{\Theta_B}{2})|0\rangle+e^{i\phi_B}\sin(\frac{\Theta_B}{2})|1\rangle\Big)_{b}, 
\end{equation}
where $\Theta_i \in [0,\pi]$ (with $i=A,B$) and $\phi_{i} \in [0,2\pi]$ are the weight and the phase of Alice's and Bob's qubits, respectively. In fact, the CNOT gate is a universal gate \cite{18}, which is a two-qubit gate that applies a Pauli-$X$ gate to the target qubit if and only if the control qubit is prepared in the state $|1\rangle$. In a generic form, one can write its effect as:
\begin{equation}
	CNOT|v,w\rangle = |v, v \oplus w\rangle.
\end{equation}
The first qubit in this representation, namely $|v\rangle$ qubit, is the control qubit. The second qubit, $|w \rangle$, is the target qubit. If the control qubit is in the state $|0\rangle$, nothing happens to the target qubit. However, if the control qubit is prepared in the state $|1\rangle$, the target qubit is flipped, i.e., if it was in state $|0\rangle$, it becomes $|1\rangle$, and vice versa.\\

The main aim of our scheme is to allow Alice to perform a non-local CNOT gate at Bob's side, where the qubit $|\Psi_A\rangle$ is the control qubit of the CNOT gate and the qubit $|\Psi_B\rangle$ is the target qubit of the CNOT gate. On the other hand, Bob could perform a non-local CNOT gate at Alice's side, where the qubit $|\Psi_B\rangle$ is the control qubit of the CNOT gate and $|\Psi_A\rangle$ is the target qubit of the CNOT gate. The CNOT is performed between Alice and Bob, by following the steps shown in the quantum circuit (Fig.\ref{FIGURE0}). 

The qubit $|Q_{aux}\rangle$ is an auxiliary qubit initialized with the state:
\begin{equation}
	|Q_{aux}\rangle = \cos(\frac{\tilde \Theta}{2})|0\rangle + e^{i\tilde{\phi}} \sin(\frac{\tilde \Theta}{2}) |1\rangle,
\end{equation}
where, $\tilde\Theta \in [0,\pi]$ and $\tilde\phi \in [0,2\pi]$ are the weight and the phase of the auxiliary qubit, respectively. Furthermore, $|R_A\rangle$ and $|R_B\rangle$ are used for the storage of projective measurement outcomes. The storage qubits $|R_A\rangle$ and $|R_B\rangle$ are prepared in the state $|0\rangle$.

\subsubsection{Performance of the protocol}
Before performing the protocol, Alice and Bob  share the Bell state:
\begin{equation}\label{5}
	|\psi^{+}_{AB}\rangle=\frac{1}{\sqrt{2}}(|00\rangle+|11\rangle)_{AB}.
\end{equation}
Where the qubits $A$ and $B$ belong to Alice and Bob, respectively. Hence, the state of the whole system is given as: 
\begin{eqnarray}
	|S_w\rangle=|Q_{aux}\rangle \otimes |\Psi_A\rangle |R_A\rangle \otimes |\Psi_B\rangle |R_B\rangle   \otimes |\psi^{+}_{AB}\rangle.
\end{eqnarray}
It is worthy to mention that the auxiliary qubit $|Q_{aux}\rangle$  is used in this scheme to select which qubit of the states $|\Psi_{A}\rangle$ and $|\Psi_{B}\rangle$, will play the role of the control qubit or the target qubit. In fact, if $|Q_{aux} \rangle = |1\rangle$, then $|\Psi_A \rangle$ is the control qubit and $|\Psi_B\rangle$  is the target qubit. Otherwise, if $|Q_{aux} \rangle= |0\rangle$, $|\Psi_B \rangle$ is the control qubit and $|\Psi_A \rangle$ is the target qubit.
\vspace{1em}
\begin{itemize}
	\item{\underline{\it{ Alice performs CNOT gate at Bob's side, namely $|Q_{aux}\rangle=|1\rangle$:}}}
\end{itemize}
\vspace{0.5em}
\begin{enumerate}
	\item{\underline{\textbf{Step One:}}} \\

	Alice performs a Toffoli gate \cite{18} using the qubits $|Q_{aux}\rangle$ and $|\Psi_A\rangle$ as a control qubits,  and the Bell qubit $|\Psi^{+}_{A}\rangle$ as the target qubit. Actually,  the Toffoli gate plays the same role as the CNOT gate if the control qubit $|Q_{aux}\rangle$  is set to the state $|1\rangle$ \cite{3}. The CNOT gate flips the target qubit if and only if the control qubit is in the state $|1\rangle$. Therefore, the total state is given as:
	\begin{eqnarray}
		|S_w\rangle^{'}&=&CNOT  \Big(|\Psi_A\rangle |R_A\rangle \otimes |\Psi_B\rangle |R_B\rangle \otimes |Q_{aux}\rangle \otimes |\Psi\rangle^{+}_{AB}\Big), \nonumber\\
		&=&\Big(\alpha\gamma|00\rangle |\Psi\rangle^{+}_{AB}+ \alpha\delta|01\rangle |\Psi\rangle^{+}_{AB}+ \beta\gamma|10\rangle |\Psi\rangle^{-}_{AB} + \beta\delta|11\rangle |\Psi\rangle^{-}_{AB}\Big) \otimes |100\rangle_{{Q_{aux}},R_{A},R_{B}}.
	\end{eqnarray}
	Whereas:
	\begin{equation*}
		\alpha=\cos(\frac{\Theta_A}{2})~;~\gamma=\cos(\frac{\Theta_B}{2})~;~\beta=e^{i\phi_A}\sin(\frac{\Theta_A}{2})~;~\delta=e^{i\phi_B}\sin(\frac{\Theta_B}{2})~;~
	\end{equation*}
	\begin{equation*}
		|\Psi\rangle^{-}_{AB}=\frac{1}{\sqrt{2}}(|10\rangle+|01\rangle)_{AB}~~ ;~~|\Psi\rangle^{+}_{AB}=\frac{1}{\sqrt{2}}(|00\rangle+|11\rangle)_{AB}
	\end{equation*}
	\item{\underline{\textbf{Step Two:}}\\

		Alice performs a Toffoli gate with the Bell qubit $|\Psi^{+}_{A}\rangle$ and $|Q_{aux}\rangle$ are the control qubits and the storage qubit, $|R_A\rangle$, as the target qubit. The state is given as follows:
		\begin{equation}
			(\alpha \gamma |00\rangle + \alpha \delta |01\rangle)\otimes \frac{1}{\sqrt{2}}(|00\rangle |0\rangle_{R_{A}} +|11\rangle |1\rangle_{R_{A}}) + (\beta\gamma|10\rangle + \beta \delta|11\rangle) \otimes \frac{1}{\sqrt{2}}(|01\rangle|0\rangle_{R_{A}} + |10\rangle|1\rangle_{R_{A}})\otimes|10\rangle_{Q{aux}, R_{B}}.
		\end{equation}
		
		\item{\underline{\textbf{Step Three:}}} \\

		Alice performs a projective measurement on the storage qubit $|R_A\rangle$ in the computational basis $\{ |0\rangle, |1\rangle \}$}. If the measurement outcome is $|1\rangle_{R_{A}}$,  the Pauli gate $\hat{\sigma}_x$ is applied on the Bell qubit $|\Psi^{+}_B\rangle$. Otherwise, if the measurement outcome is $|0\rangle_{R_{A}}$, the Pauli gate $\hat{\sigma}_{x}$ is not triggered. The \textbf{table \ref{table1}} gives the collapsed state after performing the projective measurement on the storage qubit $|R_{A}\rangle$.
	\\
	\begin{table}[ht!]
		\centering
		\begin{tabular}{ |p{3cm}||p{3cm}| |p{10cm}|}
			\hline
			\rowcolor{lightgray} Measurement result &\quad Pauli operator& \quad\quad\quad\quad\quad\quad\quad\quad Collapsed state $|S_w\rangle^{''}$ \\
			\hline
			\quad\quad$|R_{A}\rangle=|0\rangle$ & \quad\quad\quad$\hat{I}$ & $ \frac{1}{\sqrt{2}}(\alpha\gamma|00\rangle_{ab}|0\rangle_B + \alpha \delta |01\rangle_{ab} |0\rangle_B+ \beta\gamma |10\rangle_{ab} |1\rangle_B + \beta \delta |11\rangle_{ab} |1\rangle_B)$ \\
			\hline
			\quad\quad$|R_{A}\rangle=|1\rangle$ &$\quad\quad\quad\hat{\sigma_{x}}$ &  $\frac{1}{\sqrt{2}}(\alpha\gamma|00\rangle_{ab}|1\rangle_B + \alpha \delta |01\rangle_{ab} |1\rangle_B+ \beta\gamma |10\rangle_{ab} |0\rangle_B + \beta \delta |11\rangle_{ab} |0\rangle_B) $\\
			\hline
		\end{tabular}
		\caption{\label{tab:table-name} The collapsed state after performing the projective measurement on the states $|R_A\rangle$.}
		\label{table1}
	\end{table}
\\

	Similarly,  if Bob performs a projective measurement on $|R_{B}\rangle$, the collapsed state is given in \textbf{table \ref{table2}}
	\begin{table}[ht!]
		\centering
		\begin{tabular}{ |p{3cm}||p{3cm}| |p{10cm}|}
			\hline
			\rowcolor{lightgray} Measurement result & \quad Pauli operator& \quad\quad\quad\quad\quad\quad\quad\quad Collapsed state $|S_w\rangle^{''}$ \\
			\hline
			\quad\quad $|R_{B}\rangle=|0\rangle$ &\quad $\hat{I}$ & $ \frac{1}{\sqrt{2}}(\alpha\gamma|00\rangle_{ab}|0\rangle_A+ \alpha \delta |01\rangle_{ab} |0\rangle_A+ \beta\gamma |10\rangle_{ab} |1\rangle_A + \beta \delta |11\rangle_{ab} |1\rangle_A)$ \\
			\hline
			\quad\quad $|R_{B}\rangle=|1\rangle$ &\quad $\hat{\sigma_{x}}$ &  $\frac{1}{\sqrt{2}}(\alpha\gamma|00\rangle_{ab}|1\rangle_A + \alpha \delta |01\rangle_{ab} |1\rangle_A+ \beta\gamma |10\rangle_{ab} |0\rangle_A + \beta \delta |11\rangle_{ab} |0\rangle_A) $\\
			\hline
		\end{tabular}
		\caption{\label{table2} The collapsed state after performing the projective measurement on the states $|R_B\rangle$.}
		\label{tab:unitary}
	\end{table}
	\item{\underline{\textbf{Step Four:}} }\\

	A Toffoli gate is applied with the qubits $|Q_{aux}\rangle$ and $|\Psi_B\rangle$ are control qubits and $|\Psi^{+}_B\rangle$ is the target qubit. By considering the trigger qubit $|Q_{aux}\rangle =|1\rangle$, the state becomes:
	\begin{eqnarray}
		|S_w\rangle^{'''}=\frac{1}{\sqrt{2}}(\alpha\gamma|00\rangle_{ab}|0\rangle_B + \alpha \delta |01\rangle_{ab} |0\rangle_B+ \beta\gamma |11\rangle_{ab} |1\rangle_B + \beta \delta |10\rangle_{ab} |1\rangle_B).
	\end{eqnarray}
	\item{\underline{\textbf{Step Five:}}} \\

	Here, we need to get rid of  the qubit $B$, namely the Bell qubit $|\Psi^{+}_B\rangle$. We apply a Hadamard gate followed by a projective measurement in the computational basis $\{ |0\rangle, |1\rangle \}$. Depending on the measurement outcome $(|0\rangle$ or $|1\rangle)$, a Pauli gate $\hat{\sigma}_z$ is applied to the qubit $|\Psi_A\rangle$. Indeed, by applying the Hadamard gate to the Bell qubit $|\Psi^{+}_B\rangle$, the state is written as:
	\begin{eqnarray}
		H|S_w\rangle^{'''}=\frac{1}{\sqrt{2}}(\alpha\gamma|00\rangle_{ab}|+\rangle_B + \alpha \delta |01\rangle_{ab} |+\rangle_B+ \beta\gamma |11\rangle_{ab} |-\rangle_B + \beta \delta |10\rangle_{ab} |-\rangle_B),
	\end{eqnarray}
	where: 
	\begin{eqnarray}
		|\pm\rangle_B=\frac{1}{\sqrt{2}}(|0\rangle\pm|1\rangle)_B.
	\end{eqnarray}
	Now, let perform a projective measurement of the state $|\Psi^{+}_B\rangle$ in the computational basis $\{|0\rangle, |1\rangle \}$. If the measurement result is the state $|0\rangle_B$. The Pauli gate $\hat{\sigma}_z$ is not triggered. Otherwise, if the measurement result is : $|1\rangle_B$, the Pauli gate $\hat{\sigma}_z$ is applied to the state $|\Psi_a\rangle$. However, the resulting state reads: 
	\begin{eqnarray}
		\alpha\gamma |00\rangle_{ab}+ \alpha\delta |01\rangle_{ab} + \beta \gamma|11\rangle_{ab} + \beta \delta |10\rangle_{ab}.
	\end{eqnarray}
\end{enumerate}
Thus, a non-local CNOT gate is applied where the control qubit $|\Psi_A\rangle$ is at Alice's hand, and the target qubit $|\Psi_B\rangle$ is at Bob's side.		
\begin{itemize}
	\item{\underline{\it{ Alice performs CNOT gate at Bob's side, namely $|Q_{aux}\rangle=|0\rangle$:}}}
\vspace{1em}
\end{itemize}	
	
Now, by setting the auxiliary qubit $|Q_{aux}\rangle=|0\rangle$. In addition, by following the same steps as above, one can perform the CNOT gate between $|\Psi_B\rangle$ and $|\Psi_A\rangle$, where $|\Psi_B\rangle$ is the control qubit and $|\Psi_A\rangle$ is the target qubit. \\

Overall, the suggested scheme allows us to perform the dual non-local CNOT gate protocol, where the control qubit is on one side and the target qubit is on the other side. The auxiliary qubit state controls which qubit is the control  and which one is the target. Namely, if $|Q_{aux}\rangle=|1\rangle$, the qubit that is at Alice's side is the control qubit of the CNOT gate and the qubit that is at Bob's side is the target qubit. However, if $|Q_{aux}\rangle=|0\rangle$,  the qubit at Bob's side is the control qubit of the CNOT gate and Alice has the target qubit.

\subsubsection{Implementation of the protocol}
The output state  of the protocol depends on the probability of Alice and Bob via performing a projective measurements. Indeed, the probability of Alice making a projective measurement turns out to be: 
\begin{equation}
	\Pi_{A} = Tr(\rho_{aux} \rho_{a})= \frac{1}{2}+\frac{1}{2}\cos(\Theta_A)\cos(\tilde\Theta_A) + \frac{1}{2}\sin(\Theta_A)\sin(\tilde\Theta_A)Cos(\tilde\phi-\phi_{A}).
\end{equation}
Similarly, the probability of making a projective measurement by Bob reads: 
\begin{equation}
	\Pi_{B} = Tr(\rho_{aux} \rho_{b})= \frac{1}{2}+\frac{1}{2}\cos(\Theta_B)\cos(\tilde\Theta_B) + \frac{1}{2}\sin(\Theta_B)\sin(\tilde\Theta_B)\cos(\tilde\phi-\phi_{B}), 
\end{equation}
where $\rho_{aux}=|Q_{aux}\rangle \langle Q_{aux}|$ and $\rho_{i}=|\Psi_{i}\rangle \langle \Psi_{i}|$ (where $i=a,b$ is referred to Alice and Bob, respectively). Finally, by performing the protocol, Bob obtains the following state:
\begin{equation}\label{15}
	\varrho^{B}_{f}=\Pi_{A}\bar{\Pi}_{B} \omega + (1-\Pi_{A}\bar{\Pi}_{B}) \tau.
\end{equation} 
However, Alice gets the state $\varrho^{A}_{f}$ as: 
\begin{equation}\label{16}
	\varrho^{A}_{f}=\Pi_{B}\bar{\Pi}_{A} \omega^{'} + (1-\Pi_{B}\bar{\Pi}_{A}) \tau^{'}.
\end{equation} 		
The density matrices $\omega, \omega^{'}, \tau$	and $\tau^{'}$ are given in \textbf{Appendix A}.

\subsection{Efficiency of the protocol}
The concept of fidelity is a well-known quantifier which is frequently used to investigate the efficiency of the teleported states\cite{32}. In fact, it quantifies the similarity between the input and the output states. We assume that 	$\mathcal{F}(\vartheta,\varepsilon)$ is the fidelity between the states $\vartheta$ and $\varepsilon$. It is defined as \cite{33}:
\begin{equation}\label{fidelity}
	\mathcal{F}(\vartheta,\varepsilon)= ||\sqrt{\vartheta}, \sqrt{\varepsilon}||^{2}_{1}
\end{equation}
where the trace norm of an operator $J$ is given by $||J||_{1}$ =$ Tr(|J|)$, and $|J|=\sqrt{J^{\dagger} J}$. The fidelity in Eq. (\ref{fidelity}) is equal to one if and only if the states $\vartheta$  and $\varepsilon$ are similar. However, it vanishes for orthogonal states. Now, suppose that $\vartheta$ is a pure state $|\Psi\rangle$, the fidelity reduces to the following expression \cite{34}:
\begin{equation}
	\mathcal{F}(|\Psi\rangle \langle \Psi|,\varepsilon) = \langle \Psi|\varepsilon|\Psi \rangle.
\end{equation}
The above formula indicates  the probability where the state $\varepsilon$ is exactly the same state $|\Psi\rangle \langle \Psi|$. On the other hand, since the fidelity is a measure of similarity between the input and output states, then the so-called infidelity is a measure of the distinguishability between them. Therefore, it can be used to quantify the error in non-local CNOT implementation. The infidelity is defined as \cite{35}:
\begin{equation}
	\bar{\mathcal{F}}=1-\mathcal{F}(|\Psi\rangle \langle \Psi|,\varepsilon).
\end{equation}
Its worth to mention that, in our context, the infidelity $\bar{\mathcal{F}} = 1$ means the initial state is exactly the same as the final state, i.e, the non-local implementation of CNOT is fails. However, a null infidelity means that the non-local  implementation of CNOT gate is achieved optimally.
\\

Particularly, the infidelity, for the case in which  Alice has the control qubit and Bob has the target qubit of CNOT gate is given as:
\begin{equation}
	\bar{\mathcal{F}}^{A \rightarrow B}=1-\Pi_{A} \bar{\Pi}_{B}(\frac{1}{4}(1+\cos^2(\Theta_B))(1+\cos^2(\Theta_A)))-(1-\Pi_{A} \bar{\Pi}_{B})(\cos^{4}(\frac{\Theta_A}{2}) - \frac{1}{2}\sin^{2}(\Theta_B) \cos(\Theta_A)).
\end{equation}
Similarly, the expression of the infidelity where Bob  gets the control qubit reads: 
\begin{equation}
	\bar{\mathcal{F}}^{B \rightarrow A}=1-\Pi_{B} \bar{\Pi}_{A}(\frac{1}{4}(1+\cos^2(\Theta_A))(1+\cos^2(\Theta_B)))-(1-\Pi_{B} \bar{\Pi}_{A})(\cos^{4}(\frac{\Theta_B}{2}) - \frac{1}{2}\sin^{2}(\Theta_A) \cos(\Theta_B)).
\end{equation}

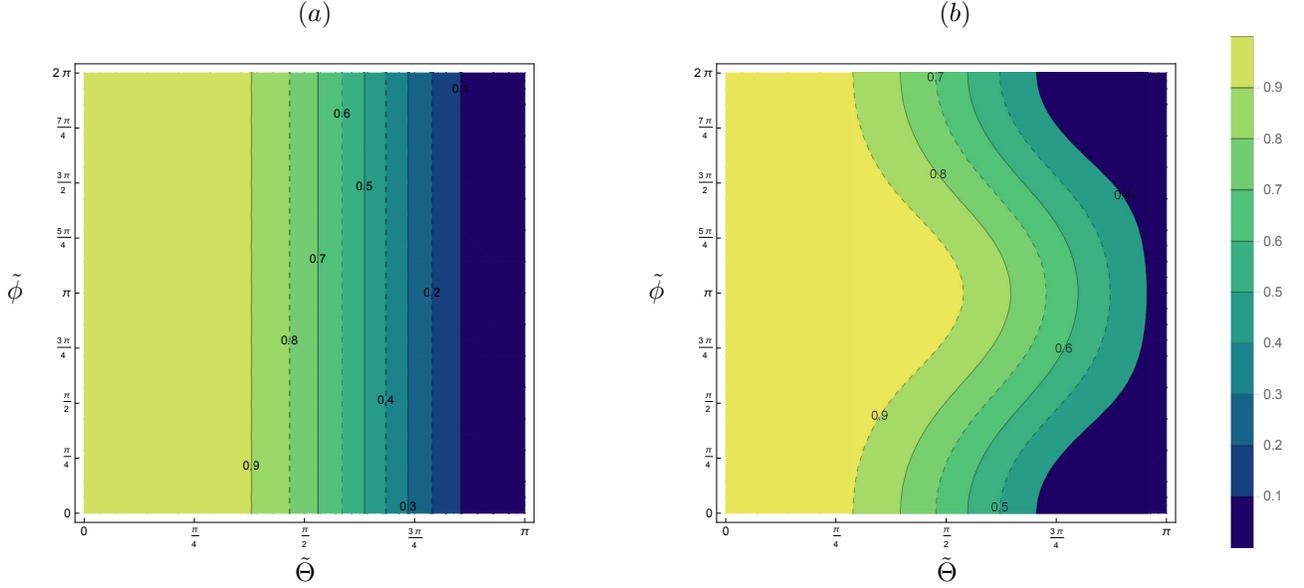
\begin{figure}[h!]
	\begin{center}
		\includegraphics[scale=0.7]{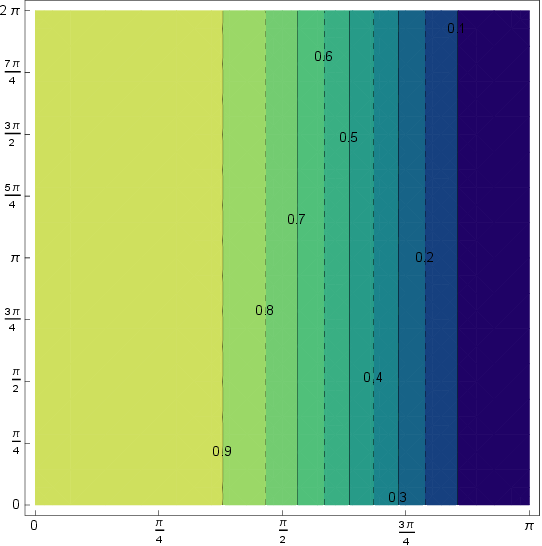}
		\put(-90,200){$(a)$}
		\put(-200,95){$\tilde{\phi}$}
		\put(-91,-12){$\tilde{\Theta}$}
		\quad\quad
		\quad\quad\quad\quad
		\includegraphics[scale=0.7]{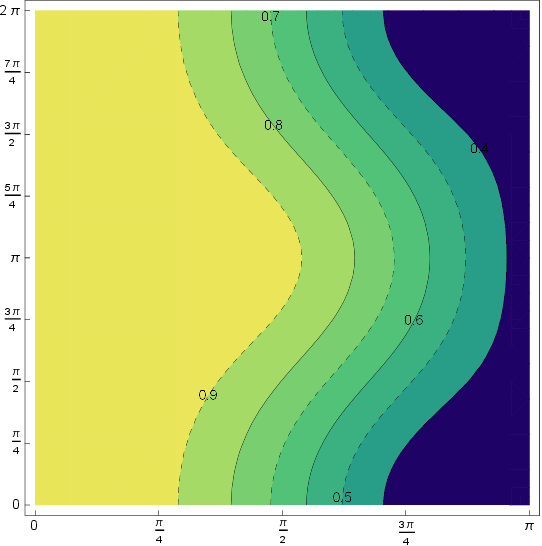}
		\put(-90,200){$(b)$}
		\put(-200,95){$\tilde{\phi}$}
		\put(-91,-12){$\tilde{\Theta}$}
		\quad\quad
		\includegraphics[scale=0.8]{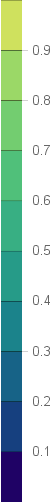}
		\caption{Infidelity (error) of the non-local CNOT implementation versus the auxiliary qubit settings, $\tilde\Theta$ and $\tilde\phi$.}
		\label{FIGURE1}
	\end{center}
\end{figure}

We examine the effect of the auxiliary qubit setting on the infidelity of the non-local implementation of the CNOT gate in Fig.(\ref{FIGURE1}). In this case, we suppose that $\Theta_A=\Theta_{B}=\pi$ and $\phi_{A}=\phi_{B}=0$, namely the initial states of Alice and Bob are prepared in classical states equivalent to $|1\rangle$. We plot the infidelity, i.e., the error of implementing the non-local CNOT gate in our case, against the auxiliary qubit settings, namely against the parameters $\tilde\Theta$ and $\tilde\phi$. In Fig.(\ref{FIGURE1}a) we assume that the control qubit is prepared in the classical state $|1\rangle$, namely that the control qubit's weight is $\Theta$ =$\pi$.  It is clear from this behavior that by increasing the weight of the auxiliary $\tilde\Theta$, the infidelity decreases gradually. Notably, for $\tilde\Theta = \pi$, the infidelity is completely vanished, which means that the non-local CNOT gate is performed with a zero rate of failure. Besides by decreasing the auxiliary qubit's weight $\tilde\Theta$, the infidelity is increased and therefore the non-local CNOT implementation fails.
In Fig (\ref{FIGURE1}b), we assume that the initial states of Alice and Bob, namely $|\Psi_{a}\rangle$ and $|\Psi_{b}\rangle$ are prepared in quantum states, where the qubits weight are prepared with $\Theta_{A} = \Theta_{B} = \frac{3\pi}{4}$. From Fig (\ref{FIGURE1}b), one can observe that by increasing the value of the auxiliary qubit weight $\tilde\Theta$, the infidelity of the non-local implementation of the CNOT gate decreases and the minimum values of the infidelity appears around $\tilde\Theta=\frac{3\pi}{4}$ for a small value of the auxiliary phase $\tilde\phi$. Furthermore, the infidelity vanishes for $\tilde\Theta=0$. It is worth remarking that while the auxiliary qubit is prepared in the classical state, namely  $|Q_{aux}\rangle = |1\rangle$, a null value of infidelity is maintained. Hence, the dual non-local CNOT implementation can be achieved optimally.

\begin{figure}[h!]
	\begin{center}
\includegraphics[scale=0.52]{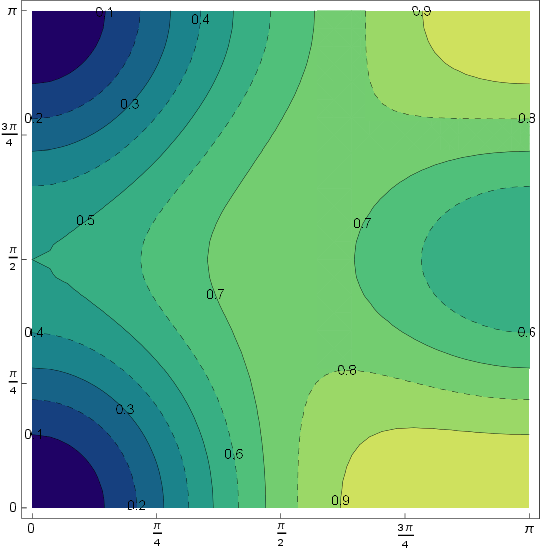}
		\put(-70,150){$(a)$}
		\put(-149,80){$\Theta_B$}
		\put(-75,-5){$\Theta_A$}
		\quad\quad\quad
		\includegraphics[scale=0.52]{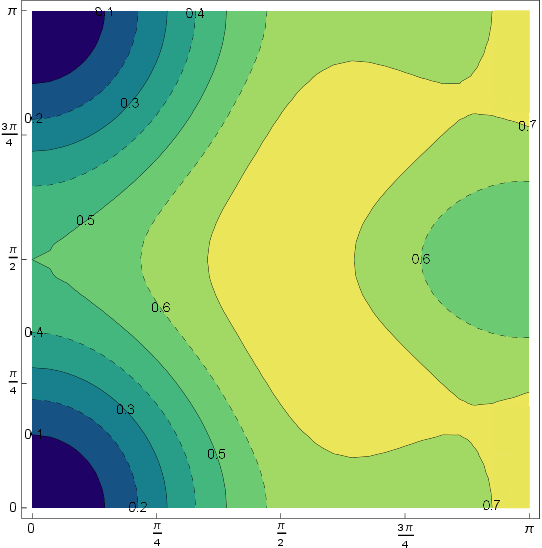}
		\put(-70,150){$(b)$}
		\put(-149,80){$\Theta_B$}
		\put(-75,-5){$\Theta_A$}\quad\quad\quad
\includegraphics[scale=0.52]{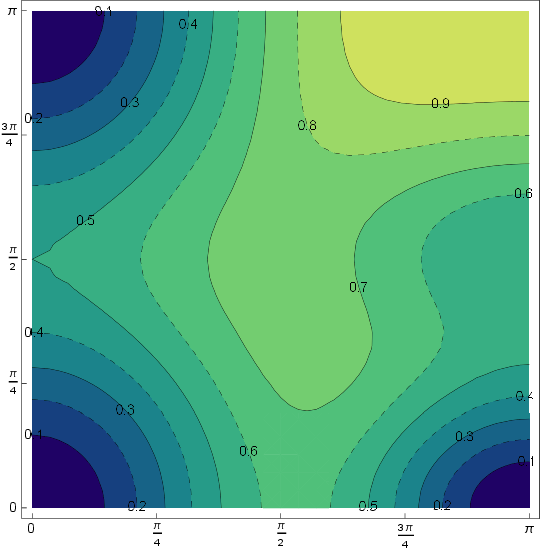}
		\put(-70,150){$(c)$}
		\put(-149,80){$\Theta_B$}
		\put(-75,-5){$\Theta_A$}\quad
\includegraphics[scale=0.6]{legend.eps}
		\caption{Infidelity of non-local CNOT gate, $\mathcal{\bar{F}}^{A \rightarrow B}$, where Alice has the control and Bob has the target qubit  against initial state settings. a): $\tilde\Theta = \pi$, b): $\tilde\Theta = \frac{\pi}{2}$ , c): $\tilde\Theta = 0$}
	\label{f4n}
\end{center}
\end{figure}

\begin{figure}[h!]
	\begin{center}

		\includegraphics[scale=0.55]{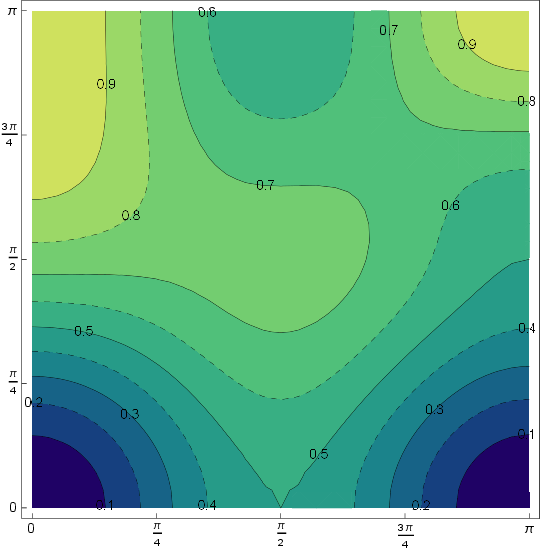}
		\put(-70,150){$(a)$}
		\put(-160,80){$\Theta_B$}
		\put(-75,-5){$\Theta_A$}\quad\quad
\includegraphics[scale=0.55]{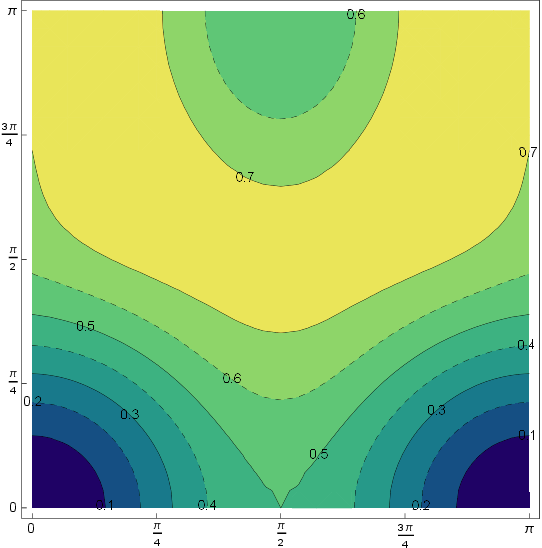}
		\put(-70,150){$(b)$}
		\put(-160,80){$\Theta_B$}
		\put(-75,-5){$\Theta_A$}
\quad\quad
\includegraphics[scale=0.55]{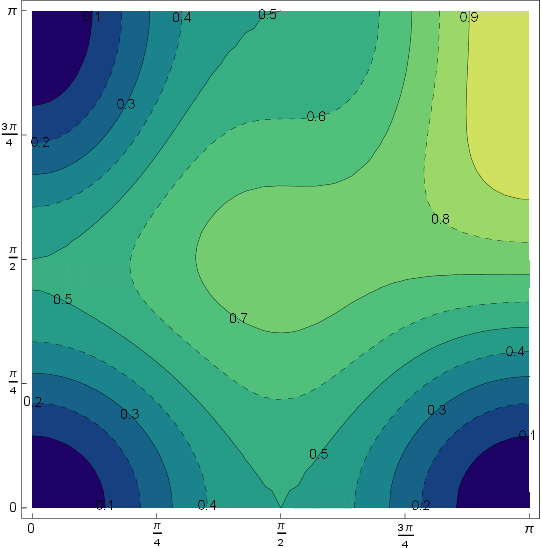}
		\put(-70,150){$(c)$}
		\put(-160,80){$\Theta_B$}
		\put(-75,-8){$\Theta_A$}
\quad
\includegraphics[scale=0.6]{legend.eps}
		\caption{Infidelity of non-local CNOT gate, $\mathcal{\bar{F}}^{B \rightarrow A}$, where Bob has the control and Alice has the target qubit  against initial state settings. a): $\tilde\Theta = \pi$, b): $\tilde\Theta = \frac{\pi}{2}$ , c): $\tilde\Theta = 0$}
		\label{f5n}
	\end{center}
\end{figure}

Next, we investigate the combined effect of the initial states settings, namely $|\Psi_{A}\rangle$, $|\Psi_{B}\rangle$ and $|Q_{aux}\rangle$, on the infidelities of the non-local CNOT gate when the control qubit at Alic's side and the target qubit is at Bob's side $\mathcal{\bar{F}}^{A \rightarrow B}$ and when the control qubit at Bob's side and the target qubit is at Alice's side $\mathcal{\bar{F}}^{B \rightarrow A}$ in Figs. (\ref{f4n}) and (\ref{f5n}), respectively. 
\\

Fig(\ref{f4n}.a) shows the infidelity behavior $\mathcal{\bar{F}}^{A \rightarrow B}$ against the initial state settings, we assume that the auxiliary qubit is prepared in the classical state $|Q_{aux}\rangle = |0\rangle$.  It is obvious from this figure that the infidelity is minimum when the control qubit weight $\Theta_{A} = 0$, namely the control state is prepared also in the classical state $|0\rangle$. Moreover, $\mathcal{\bar{F}}^{A \rightarrow B}$ increases as $\Theta_{A}$ increases. Actually, the infidelity $\mathcal{\bar{F}}^{A \rightarrow B}$ reaches the maximum bound when the control qubit and the auxiliary qubit are prepared in totally different states ($\Theta=\pi$ and $\tilde\Theta=0$). In Fig(\ref{f4n}.b), we assume that the auxiliary qubit's weight is prepared with $\tilde\Theta=\frac{\pi}{2}$. In fact, from this figure, the infidelity is minimum for $\Theta_A = \Theta_B =0$, it increases as $\Theta_A$ increases. Besides, Fig(\ref{f4n}.c) shows the infidelity $\mathcal{\bar{F}}^{A \rightarrow B}$  behavior against the initial state settings $\Theta_{A}$ and $\Theta_B$,  we assume that the auxiliary qubit's weight is prepared with $\tilde\Theta=\pi$, namely $|Q_{aux} = |1\rangle$. It is clear from this figure that the infidelity is minimum when the weight control $\Theta_{A}=0$ which means that when the initial state is prepared in the state $|0\rangle$. Moreover, the infidelity is minimum for $\Theta_{A}= \pi$, namely the control qubit is prepared in the same state as the auxiliary qubit $|\Psi_{A}\rangle =|Q_{aux}\rangle = |1\rangle$.
\\

Fig.(\ref{f5n}.a) shows the behavior of the non-local CNOT gate infidelity, whereas Bob holds the control qubit and Alice holds the target qubit of the CNOT gate, namely $\mathcal{\bar{F}}^{B \rightarrow A}$. Similarly to Fig.(\ref{f4n}.a), we assume that the auxiliary qubit weight is $\tilde\Theta = 0$. Indeed, the infidelity takes its minimum values when the control qubit weight $\Theta_{B} = 0$. In addition, the infidelity increases as $\Theta_{B}$ increases. In Fig.(\ref{f5n}.b), we set the auxiliary qubit is a superposed state, $\tilde\Theta = \frac{\pi}{2}$. Obviously, the infidelity is minimal for $\Theta_{B} = 0$ and it increases as $\Theta_{B}$ increases. However, we assume, in Fig.(\ref{f5n}.c), that the auxiliary qubit is prepared in the state $|Q_{aux}\rangle = |1\rangle$. it is clear from the plot in Fig.(\ref{f5n}.c) that the infidelity is minimum when the control qubit is prepared in the state $|\Psi_{B} = 0$, additionally, the infidelity is minimum when the control qubit and the auxliary qubit are prepapred in the same direction, namely $|\Psi\rangle_{B} = |Q_{aux}\rangle$.


\section{Comparison \& Noise analysis}
\subsection{Comparison}
In this work, a dual non-local implementation of CNOT gate is suggested where Alice can perform $CNOT$ gate at Bob's side, and also Bob can perform CNOT at Alice's side. This protocol  requires a Bell state as a quantum channel and it is based on local $CNOT$, $Toffoli$, $Hadamard$ gates, and single qubit measurement.\\

In Refs. \cite{3,30a,30,y,x} the process of CNOT teleportation needs the minimum quantum cost, namely one ebit with two classical bits. However, our suggested protocol can be performed when only one ebit is available, along with four classical bits. Nevertheless, the proposed schemes in Ref \cite{3,30a,30,y,x} are unidirectional schemes, which allow Alice only to apply the CNOT gate at Bob's side. However, our suggested scheme is bidirectional, namely that Alice can perform CNOT at Bob's qubit, and inversely, Bob can also perform the CNOT gate at Alice's qubit. In Table $(\ref{table3})$, we introduce in detail a comparison between our proposed protocol and previous ones.

\begin{table}[!ht]
	\centering
	\begin{tabular}{ |p{3cm}|p{2.5cm}|p{2.5cm}|p{3cm}|p{2cm}|}
		\hline
		\rowcolor{lightgray}~~~~Reference  &~~ Quantum cost & ~~Classical cost & ~~Scheme directions & ~~ ~~Date\\
		
		$Eisert$ $et$ $al.$  \cite{3} &  $~~~~ ~~ ~1~ebit$ &~~~~  ~$2~cbits$ & ~~~Unidirectional&  ~~ ~~(2000)\\
		\hline
		$Collins$ $et$ $al.$\cite{30a}  & $~~~~ ~~ ~1~ebit$ &~~~~  ~$2~cbits$ &~~~Unidirectional& ~~ ~~(2001) \\
		\hline
		$Huang$ $et$ $al.$\cite{30}   & $~~~~ ~~ ~1~ebit$ &~~~~  ~$2~cbits$ & ~~~Unidirectional&  ~~ ~~(2004)\\
		\hline
		$Chou$ $et$ $al.$   \cite{y}  &  $~~~~ ~~ ~1~ebit$ &~~~~  ~$2~cbits$ & ~~~Unidirectional& ~~ ~~(2018)\\
		\hline
		$Wan$ $et$ $al.$ \cite{x}  & $~~~~ ~~ ~1~ebit$ &~~~~  ~$2~cbits$ & ~~~Unidirectional& ~~ ~~(2019)\\
		\hline
		Suggested protocol& $~~~~ ~~ ~1~ebit$ &~~~~  ~$ 4~cbits$ & ~~~~Bidirectional& ~~ ~~(2023)\\
		\hline
	\end{tabular}
	\caption{Comparison between the suggested protocol and previous works. }
	\label{table3}
\end{table}

\subsection{Noise analysis}
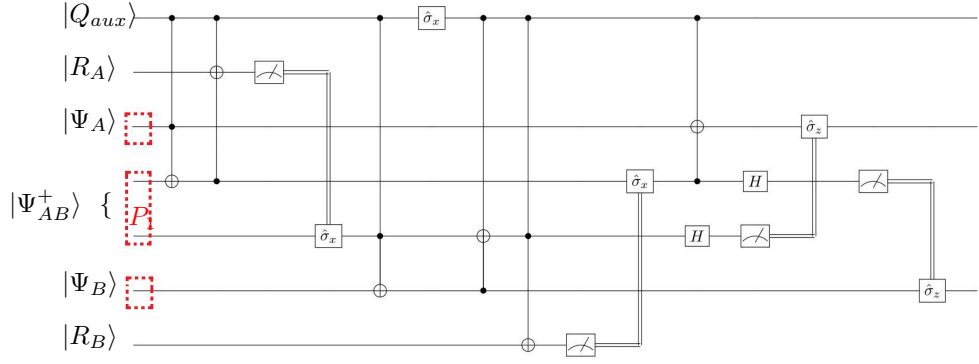
\begin{figure}[h!]
	\begin{center}
		\includegraphics[scale=.60]{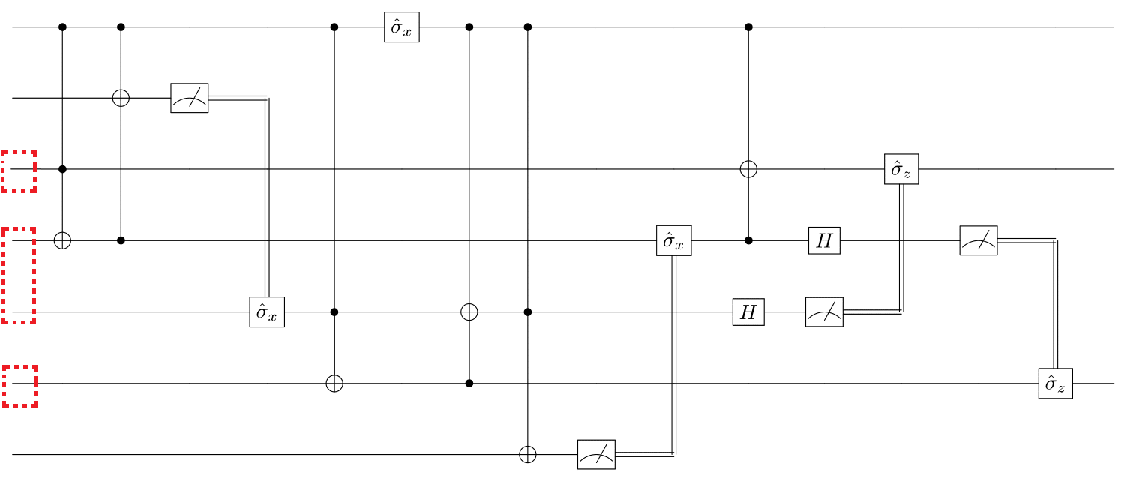}
		\put(-350,130){$|Q_{aux}\rangle$}
		\put(-350,110){$|R_A\rangle$}
		\put(-350,90){$|\Psi_A\rangle$}
		\put(-370,58){$|\Psi^{+}_{AB}\rangle~~\{$}
		\put(-325,53){$\textcolor{red}{P_1}$}
		\put(-350,28){$|\Psi_B\rangle$}
		\put(-350,08){$|R_B\rangle$}
		\caption{The circuit of dual non-local CNOT gate using Bell state. The noise, described by the dashed red box, affects the quantum channel at $P_{1}$. }
		\label{FIGUREn}
	\end{center}
\end{figure}

Let’s assume that our non-local CNOT scheme is affected by environmental noise as shown by the dashed boxes in Fig.\ref{FIGUREn}.
In this section, we investigate the effect of noise on the entangled channel $(P_1)$, by considering that the noise effect is modeled via a generalized amplitude damping channel \cite{18}. Actually, the generalized amplitude damping model $(GADM)$ is a generalization of the amplitude damping noise \cite{kunal}. The amplitude damping noise describes the physical process such as spontaneous emission or energy dissipation at zero temperature \cite{18}. However, the GADM describes the dissipation effect at a finite temperature \cite{Fujiwara}. In other words, when the temperature of the thermal bath vanishes, the GADM reduces to the amplitude damping channel. Roughly speaking, the GADM is primordial for the investigation of noise in superconducting circuits \cite{luk1}. Also, the GADM models noise in linear optical systems \cite{luk2}. Furthermore, GADM is of paramount importance in the context of secure communication \cite{luk3}.\\ 

A noisy quantum channel is defined by the completely positive trace-preserving map \cite{18}, which transforms a density operator $\rho$ into another density operator $\Phi(\rho)$ as bellow: 
\begin{equation}
	\Phi(\rho) = \sum^{3}_{i=0} \mathcal{O}_{i } \rho \mathcal{O}_{i}^{\dagger},
\end{equation}
where $\rho$ is the initial density operator, and $\mathcal{O}_{i }$ are the Kraus operators \cite{kraus}. Indeed, the ($2\times 2$) matrix representation of the Kraus operators $\mathcal{O}_k$  are given as following:
\begin{equation}
	\mathcal{O}_{0 }=\sqrt{p}
	\begin{pmatrix}
		1&0 \\
		0 & \sqrt{1-\eta}
	\end{pmatrix};\quad 
	\mathcal{O}_{1 }=\sqrt{p}
	\begin{pmatrix}
		0&\sqrt{\eta} \\
		0 & 0
	\end{pmatrix};\quad 
	\mathcal{O}_{2 }=\sqrt{1-p}
	\begin{pmatrix}
		\sqrt{1-\eta}&0 \\
		0 & 1
	\end{pmatrix};\quad 
	\mathcal{O}_{3 }=\sqrt{1-p}
	\begin{pmatrix}
		0&0 \\
		\sqrt{\eta} & 0
	\end{pmatrix},
\end{equation}
where $p\in[0,1]$ and $\eta \in [0,1]$. In addition, it is worth mentioning that for $p = 1$, the Kraus operator-sum decomposition of the AD channel can be recovered. Hence, the GADM generalizes the amplitude damping channel. Indeed, it allows the transitions from the state $|1\rangle$ to the state $|0\rangle$ for $p=1$. Inversely, it also allows the transition from the state $|0\rangle$ to the state $|1\rangle$ for $p=0$. In the former transition, the GADM acts like an amplification process. In this inspiration, the action of GADM on the teleportation quantum channel (Eq. \ref{5}), is given by applying the GADM Kraus operators on the density operator $\rho = |\Psi_{AB} ^{+}\rangle\langle_{AB} ^{+}\Psi|$. The density operator is straightforwardly obtained in the following matrix-form:  
\begin{equation}\label{noisy}
	\Phi(\rho)=\rho^{GADM}_{AB}=\frac{1}{2}
	\begin{pmatrix}
		1-\eta(1-p) & 0&0&\sqrt{1-\eta}\\
		0&\eta(1-p)&0&0\\
		0&0&\eta p &0\\
		\sqrt{1-\eta}&0&0&1- \eta p 
	\end{pmatrix},
\end{equation}

\subsection{Entanglement and implementation quantifiers}
\begin{figure}[h!]
	\begin{center}
		\includegraphics[scale=0.8]{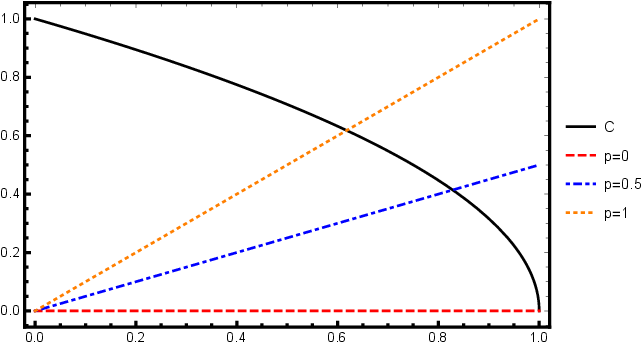}
		\put(-280,80){$\mathcal{C} \& \mathcal{\bar{F}}$}
		\put(-140,-5){$\eta$}
		\caption{Concurrence (black solid line) and the infidelity (red dashed line: $p=0$, blue dotted dashed line: $p=0.5$ and orange dotted line: $p=1$) against the GADM noise strength $\eta$.}
		\label{FIGUREe}
	\end{center}
\end{figure}

The so-called concurrence $\mathcal{C}$ is an entanglement monotone witness, widely used to quantify the amount of entanglement for bipartite quantum system $\varrho$ \cite{paper1}. Specifically, if $\mathcal{C} = 0$, then this signifies a separable state. However, if $\mathcal{C} = 1$, it indicates that the state is maximally entangled. For any bipartite state $\varrho$, the concurrence reads \cite{paper2}
\begin{equation}
	\mathcal{C} = max \{0 , \sqrt{ r_1} - \sqrt{ r_2} - \sqrt{ r_3} - \sqrt{ r_4} \}, 
\end{equation}
where $r_n$ ($n=1, 2, 3, 4$) are the eigenvalues of a matrix $R= \varrho (\sigma_{y} \otimes \sigma_{y}) \varrho^{*} (\sigma_{y} \otimes \sigma_{y})$. Whereas,  $\sigma_{y}$ is the Y-Pauli operator, and $\varrho^{*} $ denotes the complex conjugate of $\varrho$. 
\\
Now, in order to quantify the noise effect on the dual non-local CNOT implementation, we implement the suggested scheme using the noisy quantum channel given in Eq.(\ref{noisy}). In fact, the behaviors of infidelity (the implementation error) and concurrence are exhibited in Fig. (\ref{FIGUREe}).\\

Fig(\ref{FIGUREe}) displays the behavior of the entanglement's amount, measured by concurrence $\mathcal{C}$. Obviously, the entanglement's amount in the quantum channel is gradually weakening as the GADM strength $(\eta)$ increases. Moreover, it vanishes $(\mathcal{C} = 0)$ for $\eta = 1$. It is clear from Fig.(\ref{FIGUREe}) that the concurrence has the same behavior for different values of $p$. Accordingly, an optimal non local CNOT implementation, namely $\mathcal{\bar{F}}=0$, is expected for a noise-free channel $(\eta=0)$. Furthermore, the infidelity of non-local CNOT implementation increases as the noise strength $\eta$ and $p$ rise.  However, the infidelity of the non-local CNOT implementation maintain a null value when $p=0$ whatever the noise strength $\eta$ value. Indeed, for $p=0$ the quantum channel is subject to an anti-amplitude damping process that leads the quantum state of single qubit to the state $ |1\rangle \langle 1|$. Interestingly enough, in the case of $p=0$, it is obvious that the noise acts like an amplification process. Hence, the non-local CNOT operation is achieved with high efficiency even it is under high noise strength.\\
\begin{figure}[h!]
	\begin{center}
		\includegraphics[scale=0.6]{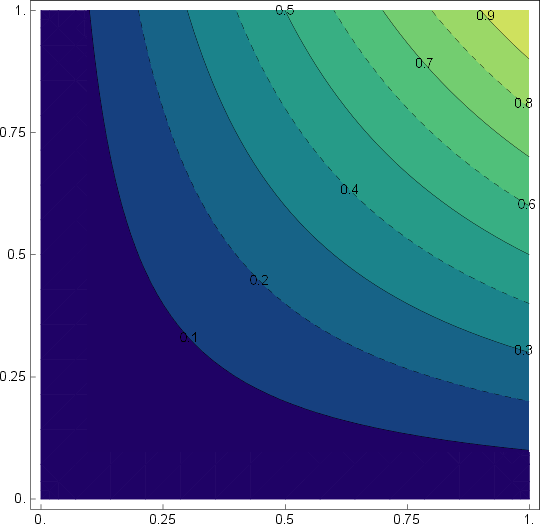}
		\put(-190,90){$p$}
		\put(-90,-10){$\eta$}
		\quad\quad
		\includegraphics[scale=.6]{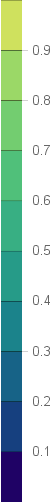}
		\caption{The infidelity versus noise strenght $\eta$ and p}
		\label{FIGUREj}
	\end{center}
\end{figure}

Fig.(\ref{FIGUREj}) displays the effect of the GADM's strength, namely $\eta$, and the effect of $p$ on the infidelity of the non-local CNOT gate. It is clear from this plot that the infidelity increases as the GADM's strength $\eta$ and $p$ increase. Indeed, the infidelity of the non-local CNOT implementation is maximal for $\eta=1$ and $p=1$, corresponding to an amplitude damping noise channel whose strength is equal to one. However, the infidelity decreases gradually as the noise's strength $\eta$ and $p$ decrease. In fact, for low values of $\eta$ and $p$, in particular when $\eta$ or $p$ does not exceed $\frac{1}{4}$, the infidelity of the non-local implementation of the CNOT gate takes a minimum bound. This result means that one can potentially implement the non-local CNOT gate process with optimal efficiency.

\newpage
\section{Conclusion}

In the realm of quantum information theory, bidirectional teleportation has become a fascinating task in recent quantum technologies, quantum metrology and quantum communication networks. Indeed, it conducts a comprehensive study to transfer the information between two partners (Alice and Bob) in two possible directions, namely from Alice to Bob and vice-versa. In this sense, we have introduced a two-way protocol for the non-local implementation of Controlled-Not gate, where two users can  perform CNOT gate bidirectionally by means of $1 ebit$ and four classical channels. In order to  implement the protocol successfully, Alice and Bob perform series of CNOT, Toffoli, Hadamard, and single-qubit measurement gates. Each user is given an arbitrary qubit, and two storage qubits which are initially prepared in a vacuum state. Any user's qubit could be the control or the trigger qubit of the CNOT gate depending on the auxiliary qubit state. \\

For the purpose of evaluating the efficiency of the suggested scheme, we have evaluated the infidelity as a measure of the distinguishably between the input state and the state after performing the CNOT operation. In fact, the infidelity can be viewed as an error in implementing the protocol. The behavior of the infidelity is investigated numerically for different initial state settings of Alice and Bob qubits. Our results showed that the maximum infidelity of the non-local implementation of the CNOT gate is obtained when the control qubit and the auxiliary qubit are polarized in opposite directions. In contrast, the minimum infidelity of the non-local CNOT implementation is obtained when the control qubit of the CNOT gate and the auxiliary qubit are polarized in the same direction.\\
We also investigated the effect of external noise on the suggested scheme by assuming that noise affect the quantum channel. We showed that by controlling noise strengths. On can achieve the dual non-local CNOT implementation optimally.
\\

In {\it conclusion}, a new dual protocol for the non-local CNOT Implementation is examined. Importantly, the efficiency of the proposed protocol depends on the initial qubit and the auxiliary qubit settings. The present results can be extended to both theoretical and experimental works. Indeed, the teleportation using the non-local implementation of the Controlled-Not Gate proposes that future studies will provide innovative methods to integrate the advantages of different methods to transmit information from one side to another and vice versa.

\renewcommand{\theequation}{A-\arabic{equation}}
\setcounter{equation}{0}  
\section*{Appendix A}

The matrices $\omega$ and $\tau$ already defined in Eqs.(\ref{15}) and (\ref{16}) are explicitly calculated as the following form: 
\begin{equation}
	\omega=
	\begin{pmatrix}
		\omega_{11} &  \omega_{12} & \omega_{13}& \omega_{14}\\
		
		\omega_{21} &  \omega_{22} & \omega_{23}& \omega_{24}\\
		
		\omega_{31} &  \omega_{32} & \omega_{33}& \omega_{34}\\
		
		\omega_{41} &  \omega_{42} & \omega_{43}& \omega_{44}
	\end{pmatrix},
\end{equation}
where, 

\begin{eqnarray}
	\omega_{11}&=& Cos^{2}(\frac{\theta_{A}}{2}) Cos^{2}(\frac{\theta_{B}}{2}), \nonumber\\
	\omega_{12}&=&e^{i\phi_{B}}Cos^{2}(\frac{\theta_{A}}{2}) Cos(\frac{\theta_{B}}{2})Sin(\frac{\theta_{B}}{2}), \nonumber\\  \omega_{13}&=&e^{i(\phi_{A}+\phi_{B})}Cos(\frac{\theta_{A}}{2}) Sin(\frac{\theta_{B}}{2})Cos(\frac{\theta_{B}}{2}) Sin(\frac{\theta_{A}}{2}), \nonumber\\
	\omega_{14}&=& e^{i\phi_{A}}Cos(\frac{\theta_{A}}{2})Cos^{2}(\frac{\theta_{B}}{2})Sin(\frac{\theta_{A}}{2}),\nonumber\\ \omega_{22}&=&Cos^{2}(\frac{\theta_{A}}{2}) Sin^{2}(\frac{\theta_{B}}{2}), \nonumber\\
	\omega_{23}&=&e^{i\phi_{A}}Cos(\frac{\theta_{A}}{2}) Sin^{2}(\frac{\theta_{B}}{2})Sin(\frac{\theta_{A}}{2}), \nonumber\\
	\omega_{24}&=&e^{-i(\phi_{B}-\phi_{A})}Cos(\frac{\theta_{A}}{2}) Sin(\frac{\theta_{B}}{2})Cos(\frac{\theta_{B}}{2})Sin(\frac{\theta_{A}}{2}), \nonumber\\
	\omega_{33}&=&Sin^{2}(\frac{\theta_{B}}{2}) Sin^{2}(\frac{\theta_{A}}{2}), \nonumber\\
	\omega_{34}&=&e^{-i\phi_{B}}Sin^{2}(\frac{\theta_A}{2})Sin(\frac{\theta_B}{2})Cos(\frac{\theta_B}{2}), \nonumber\\
	\omega_{44}&=&Sin^{2}(\frac{\theta_{A}}{2})Cos^{2}(\frac{\theta_{B}}{2}),\nonumber\\
	\omega_{21}&=&\omega_{12}^{*}, \, \omega_{31}=\omega_{13}^{*}, \, \omega_{32}=\omega_{23}^{*}, \, \omega_{41}=\omega_{14}^{*}, \, \omega_{42}=\omega_{24}^{*}, \, \omega_{43}=\omega_{34}^{*}.
\end{eqnarray}
Moreover, the density operator $\tau$ reads as: 
\begin{equation}
	\tau=
	\begin{pmatrix}
		\tau_{11} & \tau_{12} & \tau_{13}& \tau_{14}\\
		
		\tau_{21} &  \tau_{22} & \tau_{23}& \tau_{24}\\
		
		\tau_{31} &  \tau_{32} & \tau_{33}& \tau_{34}\\
		
		\tau_{41} &  \tau_{42} & \tau_{43}& \tau_{44}
	\end{pmatrix}
\end{equation}

where 

\begin{eqnarray}
	\tau_{11}&=& Cos^{2}(\frac{\theta_{A}}{2}) Cos^{2}(\frac{\theta_{B}}{2}), \nonumber\\
	\tau_{12}&=&e^{i\phi_{B}}Cos^{2}(\frac{\theta_{A}}{2}) Cos(\frac{\theta_{B}}{2})Sin(\frac{\theta_{B}}{2})\nonumber\\ \tau_{13}&=&e^{i\phi_{A}}Cos^{2}(\frac{\theta_{B}}{2}) Cos(\frac{\theta_{A}}{2})Sin(\frac{\theta_{A}}{2}), \nonumber\\
	\tau_{14}&=& Cos(\frac{\theta_{A}}{2}) Cos(\frac{\theta_{B}}{2})Sin(\frac{\theta_{A}}{2}) Sin(\frac{\theta_{B}}{2})e^{i(\phi_{A} + \phi_{B})}, \nonumber\\  \tau_{22}&=&Cos^{2}(\frac{\theta_{A}}{2}) Sin^{2}(\frac{\theta_{B}}{2}), \nonumber\\
	\tau_{23}&=& e^{-i(\phi_{B}-\phi_{A})}Cos(\frac{\theta_{A}}{2}) Sin(\frac{\theta_{B}}{2})Cos(\frac{\theta_{B}}{2}) Sin(\frac{\theta_{A}}{2}), \nonumber\\
	\tau_{24}&=&e^{i\phi_{A}}Cos(\frac{\theta_{A}}{2}) Sin(\frac{\theta_{A}}{2})Sin^{2}(\frac{\theta_{B}}{2}),\nonumber\\ 
	\tau_{33}&=&Cos^{2}(\frac{\theta_{B}}{2}) Sin^{2}(\frac{\theta_{A}}{2}), \nonumber\\
	\tau_{34}&=&e^{i\phi_{B}}Sin^{2}(\frac{\theta_A}{2})Sin(\frac{\theta_B}{2})Cos(\frac{\theta_B}{2}), \nonumber\\
	\tau_{44}&=&Sin^{2}(\frac{\theta_{A}}{2})Sin^{2}(\frac{\theta_{B}}{2}),\nonumber\\
	\tau_{21}&=&\tau_{12}^{*}, \, \tau_{31}=\tau_{13}^{*}, \, \tau_{32}=\tau_{23}^{*}, \,  \tau_{41}=\tau_{14}^{*}, \,  \tau_{42}=\tau_{24}^{*}, \, \tau_{43}=\tau_{34}^{*}
\end{eqnarray}

Similarly, the expressions of $\omega^{'}$ and $\tau^{'}$ can be straightforwardly obtained by replacing $\theta_{A}$ by $\theta_B$ and $\phi_{A}$ by $\phi_B$ in the previous matrices $\omega$ and $\tau$.

\end{document}